\documentclass[12pt]{iopart}
\usepackage{iopams}  
\usepackage[english]{babel}
\usepackage{graphicx}
\usepackage{cite}
\usepackage{bbm}

\newcommand{\bra}[1]{\ensuremath{\left\langle#1\right|}}
\newcommand{\ket}[1]{\ensuremath{\left|#1\right\rangle}}

\begin{document}

\title{Quasiprobability distribution functions from fractional Fourier transforms}

\author{Jorge A. Anaya-Contreras $^{1}$,A. Z\'u\~niga-Segundo$^1$,and  H\'ector M. Moya-Cessa$^4*$}
\address{%
$^{1}$ \quad Instituto Polit\'ecnico Nacional, ESFM, Departamento de F\'isica. Edificio 9, Unidad Profesional “Adolfo L\'opez Mateos, CP 07738 Cd. de M\'exico, Mexico\\
$^{2}$ \quad Instituto Nacional de Astrof\'{\i}sica, \'Optica y Electr\'onica, Calle Luis Enrique Erro 1, Santa Mar\'{\i}a Tonantzintla, Pue., 72840 Mexico }
\hspace{2.5cm}{$^{*}$}Correspondence: hmmc@inaoep.mx; Tel.: +52-222-266-3100

%%%%%%%%%%%%%%%%%%%%%%%%%%%%%%%%%%%%%%%%%%%%%%%%%%%%%%%%%%%%%%%%%%%%%%%%
% Abstract
%%%%%%%%%%%%%%%%%%%%%%%%%%%%%%%%%%%%%%%%%%%%%%%%%%%%%%%%%%%%%%%%%%%
%
\begin{abstract}
We show, in a formal way, how a class of complex quasiprobability
distribution functions may be introduced by using the fractional
Fourier transform. This leads to the Fresnel transform of a
characteristic function instead of the usual Fourier transform. We end the manuscript by showing a way in which the distribution we are introducing may be reconstructed by using atom-field interactions.
\end{abstract}
\begin{paragraph}
\end{paragraph}

%%%%%%%%%%%%%%%%%%%%%%%%%%%%%%%%%%%%%%%%%%%%%%%%%%%%%%%%%%%%%%%%%%%%%%%%%%%%%%%%%%%%%%%%

%\pacs{42.50.Ct, 42.50.-p, 42.50.Pq, 42.50.Dv}
%\ocis{ (050.5298) Photonic crystals;  (230.7370) Waveguides;  (350.5500) Propagation.}
%\submitto{\JPA}

\maketitle

%%%%%%%%%%%%%%%%%%%%%%%%%%%%%%%%%%%%%%%%%%%%%%%%%%%%%%%%%%%%%%%%%%%%%%%%%%%%%%%%%%%%%%%%

\section{Introduction}
It has been already shown that quasiprobability distribution
functions may be reconstructed by the measurement of atomic
properties in ion-laser interactions \cite{Leibfried} and two-level atoms
interacting with quantized fields \cite{Haroche,Davidovich}. Such measurements of the wave
function are realized usually by measuring atomic observables,
namely, the atomic inversion and polarization. 

Although the first quasiprobability distribution functions were introduced in the quantum realm \cite{Wigner,McCoy,Dirac,Kirkwood,Husimi}, they may be also used to analyze classical signals \cite{Alonso,Wolf}.

Ideal interactions, i.e., without taking into account an environment, have shown to lead to the reconstruction of the
Wigner function \cite{Davidovich} by taking advantage of its expression in terms of the parity operator. However, the interaction of a system with its
environment \cite{Amaro} leads to $s$-prametrized quasiprobability distribution
functions \cite{Royer,Wuensche,Knight}
\begin{equation} \label{one-1}
F(\alpha,s)=
\frac{2}{\pi(1-s)}\sum_{k=0}^{\infty}\left(\frac{s+1}{s-1}\right)^k
\langle k|\hat{D}^{\dagger}(\alpha)\hat{\rho} \hat{D}(\alpha)|k\rangle
\end{equation}%
where $\hat{D}(\alpha)=\exp(\alpha \hat{a}^{\dagger}-\alpha^*\hat{a})$, with $\hat{a}$ and
$\hat{a}^{\dagger}$ the annihilation and creation operators of the
harmonic oscillator, respectively, is the Glauber displacement
operator \cite{Glauber}. The state $\hat{D}(\alpha)|k\rangle=|\alpha,k\rangle$ is a so-called displaced number state \cite{Oliveira}. Note that, in order to reconstruct a given quasiprobability function it is  needed to do displace the system by an amplitude $\alpha$ and then measure the diagonal elements of the displaced density matrix.

The parameter $s$ defines different orderings and therefore different quasiprobability distribution functions (QDF). The Glauber-Sudarshan $P$-function \cite{Glauber,Sudarshan} is given for $s=1$, and is used to obtain averages of functions of {\it normal} ordered creation and annihilation operators; $s=-1$ gives the Husimi $Q$-function, used to obtain averages of functions of {\it anti-normal} ordered creation and annihilation operators, while $s=0$ is used for the {\it symmetric} ordering and gives the Wigner function.

Equation (\ref{one-1}) may be rewritten as
\begin{equation}
F(\alpha,s)=
\frac{2}{\pi(1-s)}Tr\left\{\left(\frac{s+1}{s-1}\right)^{\hat{a}^{\dagger}\hat{a}}
\hat{D}^{\dagger}(\alpha)\hat{\rho} \hat{D}(\alpha)\right\},
\end{equation}%
that, by using the commutation properties under the symbol
of trace, and the system in a {\it pure} state $|\psi\rangle$, may be casted into
\begin{eqnarray}
F(\alpha,s)&=& \frac{2}{\pi(1-s)} Tr\left\{
\hat{D}(\alpha)\left(\frac{s+1}{s-1}\right)^{\hat{a}^{\dagger}\hat{a}}
\hat{D}^{\dagger}(\alpha)\hat{\rho}\right\} \nonumber\\
&=&\frac{2}{\pi(1-s)}\langle \psi |
\hat{D}(\alpha)\left(\frac{s+1}{s-1}\right)^{\hat{a}^{\dagger}\hat{a}}
\hat{D}^{\dagger}(\alpha)|\psi\rangle.
\end{eqnarray}
Recent studies have openned the possibility of measuring, instead
of observables, non-Hermitian operators \cite{Kumar}. It would be plausible
then that such measurements could be related to complex
quasiprobability distributions like the McCoy-Kirkwood-Rihaczek-Dirac
distribution functions \cite{McCoy,Dirac,Kirkwood,Rihaczek}.

In this contribution we would like to introduce other kind of
complex quasiprobabilities that, although they could be introduced
simply by taking $s$ as a complex number, we introduce them in a
formal way by considering the fractional Fourier transform (FrFT) \cite{Namias,Agarwal,Fan}
of a signal. Then, by writing the Dirac-delta function in terms of
its FrFT, we are able to write a general expression for complex
quasiprobability distributions in terms of the Fresnel transform. 
Indeed, the representation of these complex quasiprobability
distributions in terms of a Fresnel transform implies that they
are solutions of a paraxial wave equation \cite{Davidovich}. Finally, by using an effective Hamiltonian for the atom-field interaction, we show how this quasiprobability distribution function may be reconstructed.

\section{Fractional Fourier Transform}
Up to a phase, the fractional Fourier Transform of a signal $\psi(x)$ can be written by the following expression \cite{Namias,Agarwal,Fan}
\begin{equation}
\label{TFF1}
\mathcal{F}_{\omega}\left[\psi(x)\right]= \exp \left(-i\omega\hat{a}^{\dagger}\hat{a}\right)\psi(x)\;,
\end{equation}
that may be expressed in terms of an integral transform as
\begin{equation}
\label{TFF2}
\mathcal{F}_{\omega}\left[\psi(x)\right]=\int_{-\infty}^{+\infty}dx'K(x,x';\omega)\psi(x'),
\end{equation}
where
\begin{equation}
\label{TFF3}
K(x,x';\omega)=\frac{1}{\sqrt{2\pi i}}\sqrt{\frac{e^{i\omega}}{\sin\omega}}\exp\left[i\frac{x^{2}}{2}\cot\omega+i\frac{x'^{2}}{2}\cot\omega-ixx'\csc\omega\right]\,.
\end{equation}
Then, if we consider equation (\ref{TFF3}) as a propagator,  Dirac's delta distribution function takes the form
\begin{eqnarray}
\label{TFF4}
\delta(x-x')&=&\int_{-\infty}^{+\infty}dx'' K(x,x'';-\omega)K(x'',x';\omega) \nonumber\\
&=&\frac{1}{2\pi\sin\omega}e^{i\frac{x'^{2}}{2}\cot\omega-i\frac{x^{2}}{2}\cot\omega}\int_{-\infty}^{+\infty}dx''e^{ix''(x-x')\csc\omega}\nonumber\\
&=&\frac{1}{2\pi}\exp\left[i\frac{x'^{2}}{2}\cot\omega-i\frac{x^{2}}{2}\cot\omega\right]\int_{-\infty}^{+\infty}dx'' \exp\left[ix''(x-x')\right]\,.
\end{eqnarray}
Now, if we apply the fractional Fourier transform to the Dirac delta  function we obtain
\begin{equation}
\label{TFF5}
\mathcal{F}_{\omega}\left[\delta(x-y)\right]=\int_{-\infty}^{+\infty}dx'K(x,x';\omega)\delta(x'-y)=K(x,y;\omega).
\end{equation}
Then,  applying the inverse fractional Fourier transform  to equation (\ref{TFF5}) we obtain an alternative representation of the Dirac  delta distribution function
\begin{eqnarray}
\label{TFF6}
\delta(x-y)&=&\mathcal{F}_{-\omega}\left[\mathcal{F}_{\omega}\left[\delta(x-y)\right]\right]=\int_{-\infty}^{+\infty}dx''K(x,x'';-\omega)K(x'',y;\omega)\nonumber\\
&=&\frac{1}{2\pi}\exp\left[i\frac{y^{2}}{2}\cot\omega-i\frac{x^{2}}{2}\cot\omega\right]\int_{-\infty}^{+\infty}dx' \exp\left[ix'(x-y)\right]\,.
\end{eqnarray}
From the above equation it may be  seen that there is a  phase multiplying the  usual integral representation of the Dirac delta  function, that although could be omitted by using properties of the delta function, we keep in order to obtain a quasiprobability distribution function as a fractional Fourier (Fresnel) transform of the characteristic function.

\section{Probability distribution in the phase space}
We define  $J(q,p)$ a probability distribution in the phase space as
\begin{equation}
\label{TFF7}
J(q,p)=\int_{-\infty}^{+\infty}\int_{-\infty}^{+\infty}dq'dp'\mathcal{P}(q',p')\delta(q'-q)\delta(p-p')\,,
\end{equation}
then, using equation (\ref{TFF6}), the  distribution $J(q,p)$ may be rewritten as
\begin{eqnarray}
\label{TFF8}
J(q,p)&=&\frac{1}{4\pi^{2}}e^{i\frac{q^{2}}{2}\cot\alpha-i\frac{p^{2}}{2}\cot\beta}\int_{-\infty}^{+\infty}\int_{-\infty}^{+\infty}dudv \, e^{iup-ivq}\, \times \nonumber\\
&\times& {Tr}\left\{ \hat{\rho}e^{iv\hat{q}-iu\hat{p}+i\frac{\hat{p}^{2}}{2}\cot\beta-i\frac{\hat{q}^{2}}{2}\cot\alpha}\right\} \,,
\end{eqnarray}
and because
\begin{eqnarray}
\label{TFF9}
e^{iv\hat{q}-iu\hat{p}+i\frac{\hat{p}^{2}}{2}\cot\beta-i\frac{\hat{q}^{2}}{2}\cot\alpha}&=&e^{-i\frac{u^{2}}{2}\tan\beta}e^{i\frac{v^{2}}{2}\tan\alpha}e^{iu\tan\beta\hat{q}}e^{-iv\tan\alpha\hat{p}} \times  \nonumber\\
&\times& e^{i\frac{\hat{p}^{2}}{2}\cot\beta-i\frac{\hat{q}^{2}}{2}\cot\alpha}e^{iv\tan\alpha\hat{p}}e^{-iu\tan\beta\hat{q}}\,,
\end{eqnarray}

 equation (\ref{TFF8}) takes the form
\begin{eqnarray}
\label{TFF10}
J(q,p)&=&\frac{1}{4\pi^{2}}e^{i\frac{q^{2}}{2}\cot\alpha-i\frac{p^{2}}{2}\cot\beta}\int_{-\infty}^{+\infty}\int_{-\infty}^{+\infty}dudv \,e^{iup-ivq} e^{-i\frac{u^{2}}{2}\tan\beta}e^{i\frac{v^{2}}{2}\tan\alpha} \,\times \nonumber\\ &\times& Tr\left\{ \hat{\rho}e^{iu\tan\beta\hat{q}}e^{-iv\tan\alpha\hat{p}}
e^{i\frac{\hat{p}^{2}}{2}\cot\beta-i\frac{\hat{q}^{2}}{2}\cot\alpha}e^{iv\tan\alpha\hat{p}}e^{-iu\tan\beta\hat{q}}\right\} \,,
\end{eqnarray}
that by using the equivalence 

\begin{equation}
\label{TFF11}
e^{iu\tan\beta\hat{q}}e^{-iv\tan\alpha\hat{p}}=e^{\frac{i}{2}uv\tan\alpha\tan\beta}e^{iu\tan\beta\hat{q}-iv\tan\alpha\hat{p}}\,,
\end{equation}
may be casted into the expression
\begin{eqnarray}
\label{TFF12}
J(q,p)&=&\frac{1}{4\pi^{2}}\,e^{i\frac{q^{2}}{2}\cot\alpha-i\frac{p^{2}}{2}\cot\beta}\int_{-\infty}^{+\infty}\int_{-\infty}^{+\infty}dudv \,e^{iup-ivq} e^{-i\frac{u^{2}}{2}\tan\beta}e^{i\frac{v^{2}}{2}\tan\alpha}\,\times \nonumber\\ &\times& Tr\left\{ \hat{\rho} e^{iu\tan\beta\hat{q}-iv\tan\alpha\hat{p}}e^{i\frac{\hat{p}^{2}}{2}\cot\beta-i\frac{\hat{q}^{2}}{2}\cot\alpha}e^{iv\tan\alpha\hat{p}-iu\tan\beta\hat{q}}\right\} \,.
\end{eqnarray}
\subsection{Case $\cot\alpha=-\cot\beta=\pi$}
The above quasiprobability distribution function is defined for a range of parameters $\alpha$ and $\beta$, however, for the sake of simplicity, we will consider the case $\cot\alpha=-\cot\beta=\pi$.

We may relate the quasiprobability distribution function $J(q,p)$ to the Wigner function, by noting that, for $\cot\alpha=-\cot\beta=\pi$, equation (\ref{TFF12}) has the form

\begin{eqnarray}
\label{TFF12.1}
J(q,p)&=&\frac{1}{4\pi^{2}i}e^{i\pi\left(\frac{p^{2}}{2}+\frac{q^{2}}{2}\right)}\int_{-\infty}^{+\infty}\int_{-\infty}^{+\infty}dudv \, e^{iup-ivq}e^{i\frac{u^{2}}{2\pi}+i\frac{v^{2}}{2\pi}} \times \nonumber\\
&\times& 
Tr\left\{ \hat{\rho} e^{-i\frac{u\hat{q}}{\pi}-i\frac{v\hat{p}}{\pi}}(-1)^{\hat{n}}e^{i\frac{u\hat{q}}{\pi}+i\frac{v\hat{p}}{\pi}}\right\}.
\end{eqnarray}
According to trace representation of Wigner function \cite{Knight}
\begin{equation}
\label{TFF12.2}
W\left(\frac{v}{\pi},-\frac{u}{\pi}\right)=Tr \left\{ \hat{\rho}\,\frac{1}{\pi}\, e^{-i\frac{u\hat{q}}{\pi}-i\frac{v\hat{p}}{\pi}} (-1)^{\hat{n}} e^{i\frac{u\hat{q}}{\pi}+i\frac{v\hat{p}}{\pi}} \right\},
\end{equation}
we write the distribution $J(q,p)$ as the Fresnel transform of the Wigner function
\begin{equation}
\label{TFF12.3}
J(q,p)= \frac{1}{4\pi i}e^{i\pi\left(\frac{p^{2}}{2}+\frac{q^{2}}{2}\right)}\int_{-\infty}^{+\infty}\int_{-\infty}^{+\infty}dudv \,e^{iup-ivq}e^{i\frac{u^{2}}{2\pi}+i\frac{v^{2}}{2\pi}} W\left(\frac{v}{\pi},-\frac{u}{\pi}\right)  \,.
\end{equation}

It is easy to show that the quasiprobability distribution (\ref{TFF12.3}) can be normalized 
\begin{eqnarray}
\label{TFF12.10}
\int_{-\infty}^{+\infty}\int_{-\infty}^{+\infty}dqdp\,J(q,p) \nonumber\\
=\frac{\pi}{2}\int_{-\infty}^{+\infty}\int_{-\infty}^{+\infty}dxdy \, \left[\frac{1}{2\pi}\int_{-\infty}^{+\infty}dp e^{ixp} \right] \left[\frac{1}{2\pi}\int_{-\infty}^{+\infty}dq e^{-iyq} \right]e^{-i\frac{x^{2}}{2\pi}-i\frac{y^{2}}{2\pi}} Tr \left\{ \hat{\rho} \, e^{iy\hat{q}-ix\hat{p}} \right\} \nonumber\\
= \frac{\pi}{2}\int_{-\infty}^{+\infty}\int_{-\infty}^{+\infty}dxdy \, \delta (x) \delta (y) e^{-i\frac{x^{2}}{2\pi}-i\frac{y^{2}}{2\pi}} Tr \left\{ \hat{\rho} \, e^{iy\hat{q}-ix\hat{p}} \right\} =\frac{\pi}{2} Tr\left\{ \hat{\rho} \right\} =\frac{\pi}{2} \,.
\end{eqnarray}
Therefore,  for normalization reasons, the quasiprobability distribution is finally given in the form
\begin{equation}
\label{TFF12.11}
J(q,p)=\frac{1}{4\pi^{2}}\int_{-\infty}^{+\infty}\int_{-\infty}^{+\infty}du \, dv \, e^{i u p-i v q}e^{-i\frac{u^{2}}{2\pi}-i\frac{v^{2}}{2\pi}}Tr \left\{ \hat{\rho}\,e^{i v \hat{q}-i u \hat{p}}   \right\} \,,
\end{equation}
that, by applying the change of variables $\beta = u/\sqrt{2}+iv/\sqrt{2}$  takes the form
\begin{equation}
\label{TFF12.11}
J(\alpha)=\frac{1}{2\pi^{2}}\int d^{2}\beta \, e^{\alpha \beta^{*}-\alpha^{*}\beta}e^{-\frac{i}{\pi}|\beta|^{2}}Tr \left\{ \hat{\rho}\,\hat{D}(\beta)  \right\} \,,
\end{equation}
with $\alpha = q/\sqrt{2}+ip/\sqrt{2}$.

From the above expression it is direct to show that the Wigner
function

\begin{equation}
W(\alpha)=\int d^2 \beta e^{\alpha\beta^*-\alpha^*\beta}\{\hat{\rho}
\hat{D}(\beta)\},
\end{equation}
and the function $J(\alpha)$ may be easily related by the
differential relation
\begin{equation}
J(\alpha)=\exp\left\{\frac{i}{\pi}\frac{\partial^2}{\partial
\alpha\partial \alpha^*}\right\}W(\alpha).
\end{equation}

The above quasiprobability function may be written as a trace by noting that
\begin{equation}
\label{TFF12.15}
\frac{1}{2\pi^{2}}\int d^{2}\beta\, \exp\left(-\frac{i}{\pi}|\beta|^{2}\right)\hat{D}(\beta)=\frac{1}{2i+\pi}\left(\frac{2i-\pi}{2i+\pi}\right)^{\hat{n}}
\end{equation}
that leads to the trace representation of $J(q,p)$
\begin{equation}
\label{TFF12.23}
J(q,p)=\frac{1}{2i+\pi}Tr\left\{ \hat{\rho}\,\hat{D}(\alpha)\,\left(\frac{2i-\pi}{2i+\pi}\right)^{\hat{n}} \,\hat{D}^{\dagger}(\alpha) \right\} \,.
\end{equation}
Last equation allows us to show that $J(q,p)$  is correctly normalized, for this we do the double integration
\begin{eqnarray}
\label{TFF12.24}
\int_{-\infty}^{+\infty}\int_{-\infty}^{+\infty} J(q,p)dqdp &=& Tr \left\{ \hat{\rho} \frac{2}{\pi+2i}\int \,d^{2}\alpha \hat{D}(\alpha)\hat{D}^{\dagger}\left(\alpha e^{i\theta}\right) e^{i\theta \hat{n}} \right\} \nonumber\\
 &=& Tr \left\{ \hat{\rho}\,\hat{A}\, e^{i\theta \hat{n}} \right\} \,,
\end{eqnarray}

where we have defined 
\begin{equation}
\label{TFF12.25}
e^{i\theta}=\frac{2i-\pi}{2i+\pi}\,,
\end{equation}
and
\begin{eqnarray}
\label{TFF12.26}
\hat{A}&=&\frac{2}{\pi +2i} \int d^{2}\alpha \, e^{i\sin \theta |\alpha|^{2} }\hat{D}\left( \alpha \left(1-e^{i\theta}\right) \right) \nonumber\\
&=& \frac{1}{\pi^{2}}\int\int d^{2}z_{1} d^{2}z_{2} \ket{z_{1}}\bra{z_{2}} B(z_{1},z_{2},z^{*}_{1},z^{*}_{2}) \,, 
\end{eqnarray}
with

\begin{eqnarray}
\label{TFF12.27}
B(z_{1},z_{2},z^{*}_{1},z^{*}_{2})= \frac{2}{\pi+2i} \int d^{2}\alpha  \, e^{i\sin \theta |\alpha|^{2}} \langle z_{1}|\hat{D}\left(\alpha\left(1-e^{i\theta}\right)\right)|z_{2} \rangle \nonumber\\
=\frac{2\langle z_{1}|z_{2}\rangle}{\pi+2i}   \int_{-\infty}^{+\infty}\,d\alpha_{x}  \exp\left(-\left(1-e^{i\theta}\right)\alpha_{x}^{2}+\alpha_{x}\left(\left(1-e^{i\theta}\right)z^{*}_{1}-\left(1-e^{-i\theta}\right)z_{2}\right)\right) \times
\nonumber\\ \times
\int_{-\infty}^{+\infty}\,d\alpha_{y}  \exp\left(-\left(1-e^{i\theta}\right)\alpha_{x}^{2}+i\alpha_{y}\left(\left(1-e^{i\theta}\right)z^{*}_{1}+\left(1-e^{-i\theta}\right)z_{2}\right)\right)
\nonumber\\
= \frac{2\langle z_{1}|z_{2} \rangle }{\pi+2i}\frac{\pi}{1-e^{i\theta}} \left[-\left(1-e^{-i\theta}\right)z^{*}_{1}z_{2}\right] = \left(-\frac{|z_{1}|^{2}}{2} -\frac{\left|z_{2}e^{-i\theta}\right|^{2}}{2}+z^{*}_{1}\left(e^{-i\theta}z_{2}\right)\right) \nonumber\\
=  \langle z_{1}|e^{-i\theta \hat{n}}|z_{2} \rangle \,.
\end{eqnarray}

By replacing  equation (\ref{TFF12.27}) into equation (\ref{TFF12.26}) we  obtain
\begin{equation}
\label{TFF12.28}
\hat{A}=e^{-i\theta\hat{n}}\,,
\end{equation}
that shows that equation (\ref{TFF12.24})  is correctly normalized
\begin{eqnarray}
\label{TFF12.29}
\int_{-\infty}^{+\infty}\int_{-\infty}^{+\infty} J(q,p)dqdp = Tr \left\{ \hat{\rho}\,e^{-i\theta \hat{n}}\, e^{i\theta \hat{n}} \right\} =Tr \left\{ \hat{\rho} \right\} =1 \,.
\end{eqnarray}

\section{Kirkwood distribution and  $J(q,p)$ distribution}
Being the QDF $J(q,p)$ and Kirkwood distributions complex functions we show now some differences between them.

The Kirkwood distribution is defined as \cite{Kirkwood,Rihaczek,Praxmayer1,Praxmayer2}
\begin{equation}
\label{TFF13}
K(q,p)=\frac{1}{4\pi^{2}}\int_{-\infty}^{+\infty}\int_{-\infty}^{+\infty} du \, dv \, e^{i u p - i v q}e^{i\frac{uv}{2}} Tr \left\{ \hat{\rho} e^{i v \hat{q}-i u \hat{p}} \right\}\,,
\end{equation}
or an alternative way to write it as an expectation value \cite{Moya4} is
\begin{equation}
\label{TFF14}
K(q,p)=\frac{1}{\sqrt{2}\pi}e^{\frac{q^{2}}{2}+\frac{p^{2}}{2}+iqp}\langle -i \sqrt{2}p| e^{\frac{\hat{a}^{2}}{2}}\hat{\rho}e^{-\frac{\hat{a}^{\dagger\,2}}{2}}|\sqrt{2}q \rangle.
\end{equation}

\subsection{Number state}
The Kirkwood $K(q,p)$ and $J(q,p)$ distributions for number state $\ket{n}$, are represented by the following equations
\begin{equation}
\label{TFF22}
K_{n}(q,p)=\frac{i^{n}}{2^{n}n!\pi\sqrt{2}}e^{-\frac{q^{2}}{2}-\frac{p^{2}}{2}+iqp}H_{n}(q)H_{n}(p)
\end{equation}
and
\begin{eqnarray}
\label{TFF17}
J_{n}(q,p)=\frac{1}{2i+\pi}\left(\frac{2i-\pi}{2i+\pi}\right)^{n}\exp\left(-\frac{\pi(q^{2}+p^{2})}{2i+\pi}\right)L_{n}\left(\frac{2\pi^{2}(q^{2}+p^{2})}{4+\pi^{2}}\right)\,,
\end{eqnarray}
where, $H_{n}(x)$ and $L_{n}(x)$ are  Hermite and Laguerre polynomials, respectively.
\subsection{Superposition of two coherent states}
Now, we consider a superposition of two coherent states as:
\begin{equation}
\label{TFF23}
\ket{\psi_{\pm}}  =\frac{1}{\sqrt{2\pm2\mathrm{Re}\langle\alpha_{1}|\alpha_{2}\rangle}}\left(\ket{\alpha_{1}}\pm\ket{\alpha_{2}}\right)\,,
\end{equation}

\begin{equation}
\label{TFF23}
\ket{\psi_{\pm}}=\frac{1}{\sqrt{2\pm2\mathrm{Re}\langle \alpha_{1}|\alpha_{2}\rangle}}\left(\ket{\alpha_{1}}\pm\ket{\alpha_{2}}\right)\,,
\end{equation}
where $\alpha_{k}=q_{k}/\sqrt{2}+ip_{k}/\sqrt{2}$, such that the Kirkwood $K(q,p)$ and the $J(q,p)$ distributions for the superposition of two coherent states, 
$\ket{\psi_{\pm}}$, is given by

\begin{eqnarray}
\label{TFF25}
J_{\pm}(q,p)= \frac{1}{2i+\pi}\frac{1}{2\pm2\mathrm{Re}\langle \alpha_{1}|\alpha_{2}\rangle }\left(\exp\left(-\frac{\pi}{2i+\pi}\left((q-q_{1})^{2}+(p-p_{1})^{2}\right)\right)\right)\nonumber\\
\frac{1}{2i+\pi}\frac{1}{2\pm2\mathrm{Re}\langle \alpha_{1}|\alpha_{2}\rangle}\left( \exp\left(-\frac{\pi}{2i+\pi}\left((q-q_{2})^{2}+(p-p_{2})^{2}\right)\right)\right)\nonumber\\
\pm\frac{1}{2i+\pi}\frac{1}{2\pm2\mathrm{Re}\langle \alpha_{1}|\alpha_{2}\rangle }\left( \exp\left(\frac{i}{2}\left(q\left(p_{1}-p_{2}\right)-p\left(q_{1}-q_{2}\right)\right)\right)\langle\alpha_{2}-\alpha|e^{i\theta}(\alpha_{1}-\alpha)\rangle\right)\nonumber\\
\pm\frac{1}{2i+\pi}\frac{1}{2\pm2\mathrm{Re}\langle \alpha_{1}|\alpha_{2}\rangle}\left( \exp\left(-\frac{i}{2}\left(q\left(p_{1}-p_{2}\right)-p\left(q_{1}-q_{2}\right)\right)\right)\langle\alpha_{1}-\alpha|e^{i\theta}(\alpha_{2}-\alpha)\rangle\right) \nonumber\\
\end{eqnarray}

and
\begin{eqnarray}
\label{TFF27}
K_{\pm}(q,p)=\frac{1}{\sqrt{2}\pi}\frac{\exp\left(-\frac{q^{2}}{2}-\frac{p^{2}}{2}+iqp\right)}{2\pm2\mathrm{Re}\langle \alpha_{1}|\alpha_{2}\rangle}\exp\left(-\frac{1}{2}(q^{2}_{1}+p^{2}_{1}) +(qq_{1}-pp_{1})\right) \times \nonumber\\
\times \exp\left(\frac{i}{2}q_{1}(p_{1}+2p)+\frac{i}{2}p_{1}(q_{1}-2q)\right)\nonumber\\
+
 \frac{1}{\sqrt{2}\pi}\frac{\exp\left(-\frac{q^{2}}{2}-\frac{p^{2}}{2}+iqp\right)}{2\pm2\mathrm{Re}\langle \alpha_{1}|\alpha_{2}\rangle}\exp\left(-\frac{1}{2}(q^{2}_{2}+p^{2}_{2}) +(qq_{2}-pp_{2})\right)\times \nonumber\\
\times \exp\left(\frac{i}{2}q_{2}(p_{2}+2p)+\frac{i}{2}p_{2}(q_{2}-2q)\right)\nonumber\\
\pm
 \frac{1}{\sqrt{2}\pi}\frac{\exp\left(-\frac{q^{2}}{2}-\frac{p^{2}}{2}+iqp\right)}{2\pm2\mathrm{Re}\langle \alpha_{1}|\alpha_{2}\rangle}\exp\left(-\frac{1}{2}(q^{2}_{1}+p^{2}_{2}) +(qq_{1}-pp_{2})\right)\times \nonumber\\
\times \exp\left(\frac{i}{2}q_{2}(p_{2}+2p)+\frac{i}{2}p_{1}(q_{1}-2q)\right)\nonumber\\
\pm
\frac{1}{\sqrt{2}\pi}\frac{\exp\left(-\frac{q^{2}}{2}-\frac{p^{2}}{2}+iqp\right)}{2\pm2\mathrm{Re}\langle \alpha_{1}|\alpha_{2}\rangle}\exp\left(-\frac{1}{2}(q^{2}_{2}+p^{2}_{1}) +(qq_{2}-pp_{1})\right)\times \nonumber\\
\times \exp\left(\frac{i}{2}q_{1}(p_{1}+2p)+\frac{i}{2}p_{2}(q_{2}-2q)\right)\,,
\end{eqnarray}
respectively.
\begin{figure}[h]
\centering
\includegraphics[width=8cm]{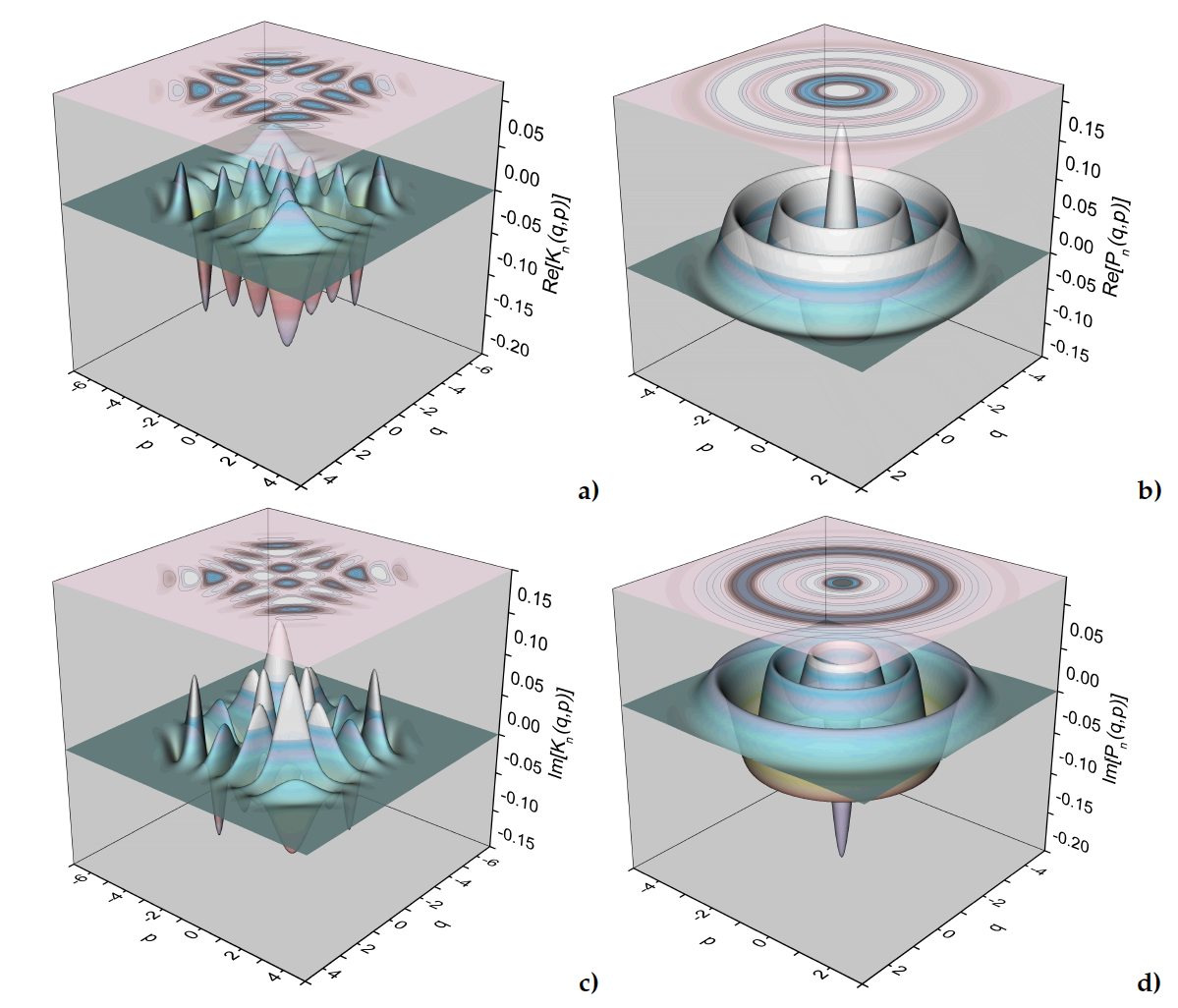}
\caption{In figures a) and c) we can see the phase space distribution of the real  and imaginary parts of the Kirkwood  function for a  number state $\ket{n=3}$. In figures b) and d) we see the distribution  $J(q,p)$, for the same number state, again, the real and imaginary parts, respectively.}
\end{figure}
We plot both distribution in Figures 1 and 2. In both figures a more uniform behaviour may be seen in the QDF $J_{\pm}(q,p)$ than in the Kirkwood function. In fact, the real and imaginary parts of the distribution we have introduced here, look like Wigner function for number states (Fig. 1)  and Scrh\"odinger cat states (Fig. 2).
\begin{figure}[h]
\centering
\includegraphics[width=8cm]{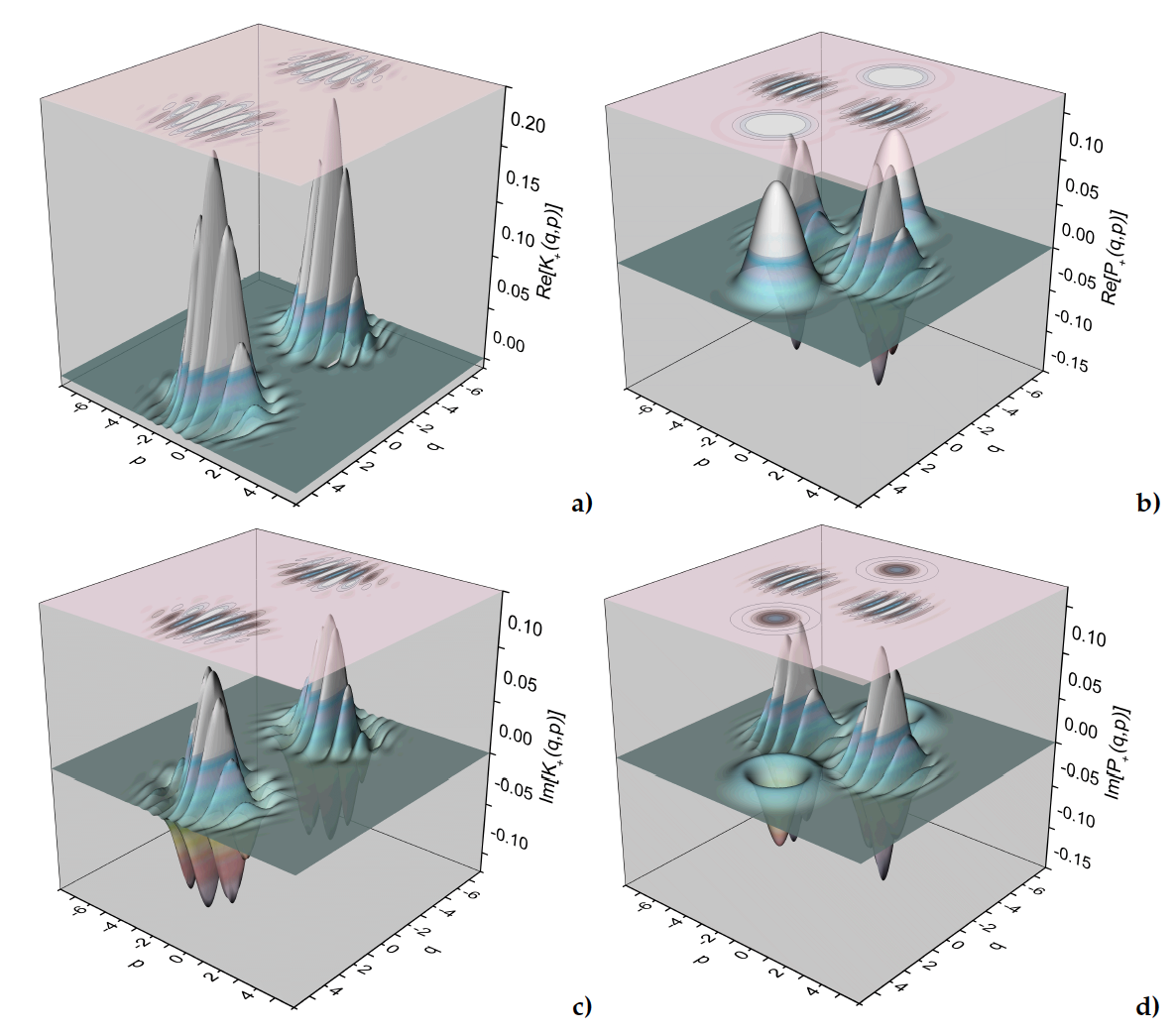}
\caption{In figures a) and c) we can see the phase space distribution of the real  and imaginary parts of the Kirkwood  function for two superposition of coherent states $\ket{\psi_{+}}$ wiht $q_{1}=-q_{2}=4$ and $p_{1}=p_{2}=0$. In figures b) and d) we see the distribution  $J(q,p)$, again, the real and imaginary parts, respectively.}
\end{figure}
\section{Reconstruction of distribution $J(\alpha)$}

It is not difficult to show that the real part of QDF $J(\alpha)$ may be measured. This can be achieved by measuring the atomic polarization in the dispersive interaction
between an atom and a quantized field \cite{Davidovich}, whose
Hamiltonian reads
\begin{equation}
\hat{H} =-\chi \hat{a}^{\dagger}\hat{a}\hat{\sigma}_z,
\end{equation}%
with $\hat{\sigma}_z=|e\rangle\langle e|-|g\rangle\langle g |$, the
Pauli matrix corresponding to the atomic inversion operator, where
$|g\rangle$ and $|e\rangle$ represent the ground and excited
states of the two-level atom. The parameter
 $\chi$ is the dispersive coupling constant. The above Hamiltonian yields the evolution operator
\begin{equation}
\hat{U}(t)=\exp\{-i\chi t \hat{a}^{\dagger}\hat{a}\hat{\sigma}_z\},
\end{equation}
from where we can obtain the evolved wavefunction $|\psi(t)\rangle=\hat{U}(t)|\psi(0)\rangle$, that allows the calculation of averages of different observables. 

The average of observable $\hat{\sigma}_x=|e\rangle\langle
g|+|g\rangle\langle e |$ then can be obtained for an arbitrary
initial field, which we conveniently write as
$|\psi_F(0)\rangle=\hat{D}^{\dagger}(\alpha)|\phi(0)\rangle$ and the atom is
initially in a superposition of atomic states, $|\psi_A(0)\rangle=
\frac{1}{\sqrt{2}}(|g\rangle+|e\rangle)$. Then we write

\begin{equation}
\langle \hat{\sigma}_x(t)\rangle=\frac{1}{2}\left(\langle
\phi(0)|\hat{D}(\alpha)\exp\{2i\chi t \hat{a}^{\dagger}\hat{a}\}\hat{D}^{\dagger}(\alpha)|\phi(0)\rangle+c.c.\right).
\end{equation}

It is also easy to show that the imaginary part of the QDF may be
associated to the observable  $\hat{\sigma}_y=i(|e\rangle\langle
g|-|g\rangle\langle e |)$
\begin{equation}
\langle\hat{\sigma}_y(t)\rangle=\frac{i}{2}\left(\langle
\phi(0)|\hat{D}(\alpha)\exp\{2i\chi t a^{\dagger}a\}\hat{D}^{\dagger}(\alpha)|\phi(0)\rangle-c.c.\right).
\end{equation}
If we set the interaction time $t=\frac{\arctan \frac{4\pi}{\pi^2-4}}{2\chi}$, we obtain that
\begin{equation}
Re\{J(\alpha)\}\propto \langle\hat{\sigma}_x\rangle, \qquad Im\{J(\alpha)\}\propto \langle\hat{\sigma}_y\rangle.
\end{equation}
Therefore, by measuring the polarizations $\hat{\sigma}_x$ and $\hat{\sigma}_y$ we are able to measure the QDF $J(\alpha)$.
\section{Conclusions}

We have introduced a set of parametrized (in terms of $\alpha $ and $\beta$) quasiprobability
distribution functions, equation (\ref{TFF12}), by using the fractional Fourier transform.
This has lead us to generalize QDF  to Fresnel transforms of the characteristic function instead of their usual Fourier transforms. We
have also shown how such QDF may be recosntructed in the dispersive
atom-field interaction. We have also given a (differential)
relation that allows the calculation of the newly introduced QDF
from the Wigner function. 
\\
\\
{\bf Acknowledgments}: We thank CONACYT for support.
\\

\end{document}